\documentclass[conference]{IEEEtran}
\IEEEoverridecommandlockouts
\usepackage{cite}
\usepackage{amsmath,amssymb,amsfonts}
\usepackage{algorithmic}
\usepackage{graphicx}
\usepackage{textcomp}
\usepackage{xcolor}
\usepackage{url,times}
\usepackage{color}
\usepackage{hyperref}
\usepackage{multirow, boldline}
\usepackage{xcolor, soul}

\usepackage{subcaption} 

\usepackage{pdfpages}   

\usepackage[export]{adjustbox}

\def\BibTeX{{\rm B\kern-.05em{\sc i\kern-.025em b}\kern-.08em
    T\kern-.1667em\lower.7ex\hbox{E}\kern-.125emX}}
\begin{document}


\title{A Comparative Study on 1.5T - 3T MRI Conversion through Deep Neural Network Models}

\author{ \IEEEauthorblockN{Binhua
    Liao\IEEEauthorrefmark{1}\IEEEauthorrefmark{2},
    Yani Chen\IEEEauthorrefmark{1},
    Zhewei Wang\IEEEauthorrefmark{1},
    Charles D. Smith\IEEEauthorrefmark{3}, Jundong
    Liu\IEEEauthorrefmark{1}}
  \\
  \IEEEauthorblockA{\IEEEauthorrefmark{1}School of Electrical
    Engineering and Computer Science, Ohio University, USA}
  \IEEEauthorblockA{\IEEEauthorrefmark{2}College of Mathematics and
    Statistics, Huazhong Normal University, PR China}
  \IEEEauthorblockA{\IEEEauthorrefmark{3}Department of Neurology,
    University of Kentucky, USA}}

\maketitle

\begin{abstract}
  In this paper, we explore the capabilities of a number of deep
  neural network models in generating whole-brain 3T-like MR images
  from clinical 1.5T MRIs. The models include a fully convolutional
  network (FCN) method and three state-of-the-art super-resolution
  solutions, ESPCN \cite{shi2016subpix}, SRGAN \cite{Jun-Yan2017Photo}
  and PRSR \cite{Ryan2017Pixel}. The FCN solution, U-Convert-Net,
  carries out mapping of 1.5T-to-3T slices through a U-Net-like
  architecture, with 3D neighborhood information integrated through a
  multi-view ensemble. The pros and cons of the models, as well the
  associated evaluation metrics, are measured with experiments and
  discussed in depth.  To the best of our knowledge, this study is the
  first work to evaluate multiple deep learning solutions for
  whole-brain MRI conversion, as well as the first attempt to utilize
  FCN/U-Net-like structure for this purpose.

\end{abstract}


\vspace{0.1in}


%
\IEEEpeerreviewmaketitle

\begin{figure*}
\centering
\includegraphics[height=3.7in]{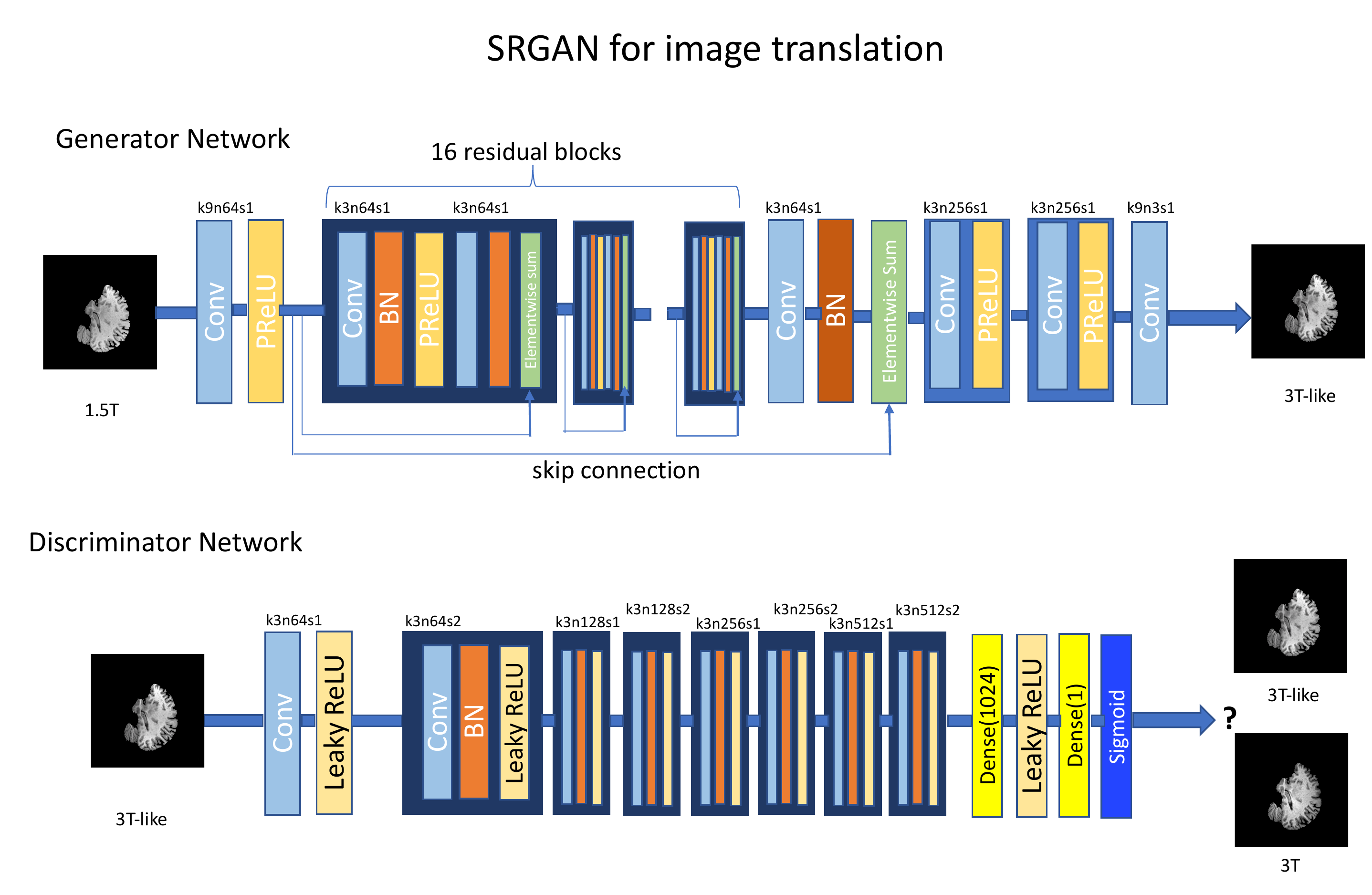} 
\caption{The overall architecture of SRGAN for MRI conversion:
  generator and discriminator with corresponding kernel size (k),
  number of feature maps (n) and stride (s) indicated for each
  convolutional layer.}
 \label{fig:srganmode}
 \end{figure*}

\section{Introduction}

Magnetic resonance imaging (MRI) is widely used in neuroimaging and the
popularity is due to its non-invasive nature, high soft tissue
contrast, as well as the availability of safe intracellular contrast
agents.
Currently 1.5 tesla (T) short-bore MRI is the standard technology for
clinical use. However, 3T (and even 7T) MRI scanners are becoming
increasingly more desirable, as they can provide extremely clear and
vivid images. Comparing with 1.5T, 3T MR images have higher
signal-to-noise ratios (SNR) and higher contrast-to-noise ratios (CNR)
between gray and white matter. The latter make 3T MRI a better choice
for brain tissue segmentation, as well as a generally preferred
modality in neuroimaging studies.




While the availability of 3T MRI has increased significantly over the
past decade, the majority of clinical scanners across the US are still
1.5T systems. Converting 1.5T images into 3T-like images, if with
great fidelity, would help physicians make better informed diagnosis
and treatment decisions. In addition, historical 1.5T MR images in
various ongoing longitudinal studies can also be brought into a better
use. One of such examples is the Alzheimer's Disease Neuroimaging
Initiative (ADNI) project -- 1.5T was the major MRI modality in ADNI
1, the first stage of the project, but the acquisition switched to 3T
alone in later stages (ADNI GO, 2 and 3).
Converting 1.5T images into 3T-like counterparts may allow the
datasets generated in such studies to be delivered in a more uniform
form.

Establishing a nonlinear spatially-varying intensity mapping between
two images is a challenging task. The efforts to tackle this problem
can trace back to at least the {\it Image Analogies} model
\cite{hertzmann2001image}, which relies on a nonparametric texture
model \cite{efros1999texture} to learn the mapping on a single pair
of input-output images. The emerge of the powerful deep learning
paradigm in recent years makes the task more viable.  {\it Generative
  adversarial networks} (GAN) \cite{isola2016image, Wolterink2017Deep,
  Zili2017DualGAN, Jun-Yan2017Unpaired,Jun-Yan2017Photo} and
pixel-RNN/CNN \cite{Ryan2017Pixel} are among the models that have been
applied for modality conversion, producing impressive results.


The original GAN model by Goodfellow {\it et al.}
\cite{goodfellow2014generative} was designed to generate images that
are similar to the training samples. Several later solutions,
including DualGAN \cite{Zili2017DualGAN}, CycleGAN
\cite{Jun-Yan2017Unpaired} and DiscoGAN \cite{DiscoGAN}, take the
similar idea to train image-to-image translation with unpaired natural
images.
The CycleGAN model has been
adopted 
to synthesize CT images from MRIs \cite{Wolterink2017Deep}.  While
flexible and with broad applicability, this group of solutions reply
on the distribution of real samples instead of paired inputs/outputs,
even if the latter are available. Consequently, the results from this
group can be rather unstable and far from uniformly positive
\cite{Jun-Yan2017Unpaired}. Some GANs, including pix2pix
\cite{isola2016image} and PAN \cite{PAN}, take paired training samples
to trade flexibility for stability.


With paired input/output samples, MR modality conversion could be
implemented as a special case of super-resolution, where one or
multiple low-resolution images are combined to generate images with
higher spatial resolution. Traditional super-resolution solutions
include reconstruction-based methods
\cite{van2012super,shilling2009super,shi2015lrtv,woo2012super}, and
example-based methods
\cite{yang2012coupled,gao2012image,zhang2012hierarchical,manjon2010mri,roy2013magnetic,huynh2016estimating,burgos2014attenuation}. Under
the deep learning framework, numerous new super-resolution solutions
have recently been developed, in both the computer vision
\cite{shi2016subpix, Ryan2017Pixel} and medical image computing
\cite{bahrami2016reconstruction}, \cite{Bahrami20167t},
\cite{deoni2022simultaneous}, \cite{zhao2020accurate, liu2018fusing,
  srinivasan2021realizing} communities.
SRGAN \cite{Jun-Yan2017Photo}, a model designed to recover the finer
texture details even with large upscaling factors, is commonly
regarded as one of the state-of-the-art solutions.


Fully convolutional networks (FCN) proposed by Long {\it et al.}
\cite{long2015fully} was primarily designed for image segmentation,
which can also be regarded as a special type of modality mapping --
from gray-valued intensities to binary-valued labels. U-Net
\cite{ronneberger2015u} and its variants \cite{chen2017hippocampus,
  chen2017accurate,wang2018ensemble,wang2019residual,
  sun2019tracecaps} follow the similar idea of FCN and rely on skip
connections to
concatenate features from the contracting (convolution) and expanding
(deconvolution) paths. In theory, an FCN with a proper setup can
potentially describe any intensity mapping between two
modalities. However, such capacity of FCN has yet been explored for
general-purpose modality conversion. It should be noted that, Nie {\it
  et al.}  \cite{Nie2016fcn} use a convolutional network for MR to CT
conversion, but their network structure is not FCN/U-Net equivalent,
as no pooling, skip connections, contracting and expanding components
are utilized.

In this paper, we explore the capability of a number of
super-resolution (SR) and segmentation models in handling modality
conversion. More specifically, we adopt SR models including ESPCN
\cite{shi2016subpix}, SRGAN and PRSR \cite{Ryan2017Pixel}, and modify
Chen's segmentation model \cite{chen2017hippocampus}, to convert 1.5T
whole-brain MR images into 3T. Experiments are conducted with ADNI
data. To the best of our knowledge, this study is the first work to
compare and evaluate multiple deep learning solutions, based on
various performance metrics, for whole-brain MRI conversion.

\section{Mothod}

The MR conversion models to be analyzed in this study are modified
from SR and segmentation solutions, respectively.  In this section, we
introduce them in detail.

\subsection{Modified from Super-Resolution Solutions}


{\bf SRGAN}
is designed to generate 4$\times$upscaled photo-realistic natural
images with highly perceptual quality. Focused on recovering finer
texture details in up-scaled images, SRGAN adopts a perceptual loss
function that consists of an adversarial loss and a content loss. As a
super-resolution solution, SRGAN produces outputs that have different
sizes from inputs. To suit for our MRI conversion task, we remove one
upsampling layer to make the input/output of equal size.

As shown in Fig.~\ref{fig:srganmode}, the modified SRGAN for MR
conversion model consists of two major components: generator and
discriminator. The generator part is a deep residual network (ResNet)
with skip-connections, generating 3T-like images from 1.5T inputs. The
goal of the generator is to be able to produce 3T images so realistic
that it would be able to fool the discriminator. The discriminator, on
the other hand, is configured as a classification CNN and its goal is
to be trained as sharp as possible to distinguish fake 3T from real 3T
images. With this setup, the generator can eventually learn to create
outputs that are highly similar to real 3T images.


{\bf ESPCN}
uses two convolutional layers to extract feature maps from low
resolution image and then applies a sub-pixel convolution layer to
transform these feature maps back to an enlarged super resolution
image. The sub-pixel layer is designed to be very efficient, which
reduces the computational complexity of the model and enables the
system to achieve real-time super-resolution of 1080p videos on a
single K2 GPU.
                
{\bf PRSR} 
is a super resolution model build upon ResNet and PixelCNNs (a
probabilistic generative model) that is capable of enlarging small
input image to a wide range of plausible high-resolution images with
large amplification factors. Experiments show that the transformed
images obtain high rate of perceptual evaluation by humans.

\begin{figure}  
\centering
\includegraphics[height=3.5in]{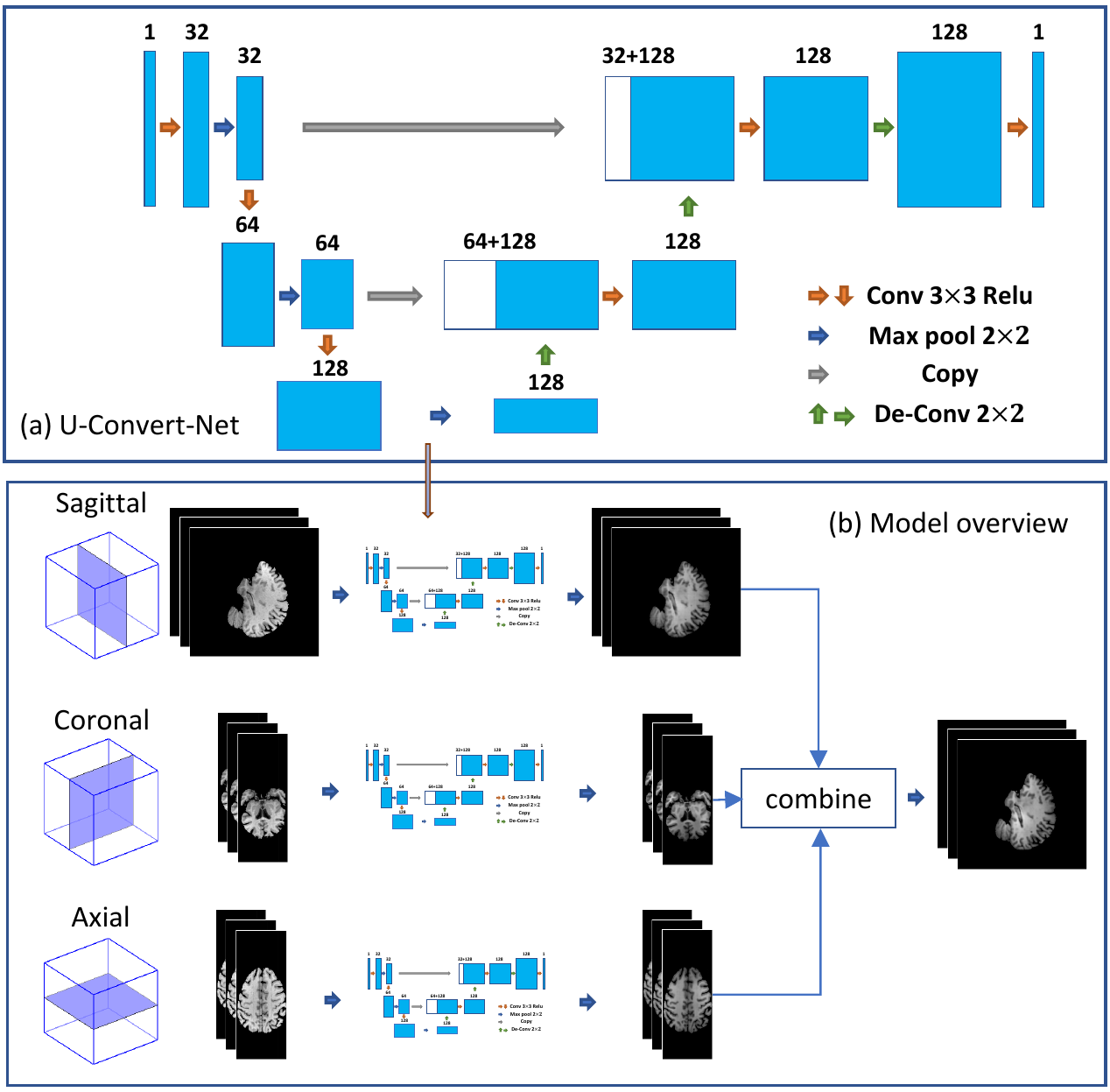} 
  \vspace{-0.15in}
  \caption{\small The overall architecture of Multi-view U-Convert-Net: (a)
    U-Convert-Net and (b) Model overview.}
 \label{fig:mainmode}
 \end{figure}

 \subsection{Modified from FCN-based Segmentation Solution}

 As shown in Fig.~\ref{fig:mainmode}, the segmentation-based model
 consists of two major components. First, U-Convert-Net, a U-Net-like
 network, is trained to convert two-dimensional (2D) slices extracted
 from
 1.5T MR volumes into slices of 3T MR images. Second, the converted 3T
 slices from three different views are fused together to generate the
 final 3T MR volumes with enhanced structural details and intensity
 contrasts.

 The main architecture of U-Convert-Net is modified from U-Net
 \cite{ronneberger2015u}.  Similar to U-Net, our U-Convert-Net (shown
 in Fig.~\ref{fig:mainmode}(a)) is also comprised of encoding and
 decoding paths. 1.5T MR slices are inputs and go through several
 ``convolution + pooling'' layers to be encoded into high-level latent
 features. These features will then be sent through several 
 ``deconvolution/upsampling'' layers to reconstruct the target 3T MR
 images. Structural information lost during the pooling procedure
 would be added back through ``bridges'': latent features in the
 encoded path are concatenated with the corresponding feature maps
 from the decoding path.
 
 U-Convert-Net is different from U-Net 
 in several aspects.
 The first major alternation is to the loss function.  U-Net
 uses cross-entropy as their loss, which is a reasonable
 choice to seek for intensity-to-label mappings. For
 intensity-to-intensity translation, mean-square-error (MSE) would be
 more appropriate, as any deviation from the target 3T image should be
 penalized. 

 The second set of modifications are made in order to reduce the
 training time and make the structure reusable for different views,
 whose input image dimensions may be different.  We use padded
 convolution to maintain the same spatial dimension of the data flow
 throughout the convolutional layers.  This is in contrast with U-Net,
 where the images are altered at input to accommodate dimension
 decrement along convolutions. We only keep one convolutional layer
 prior to each pooling layer to reduce the number of parameters of the
 network. The size of the filters is kept as $3\times3$. Third, to
 ease the training procedure, explicit data augmentation as in U-Net,
 which would increase training time, is replaced with dropout
 operations on several deconvolution layers to prevent
 overfitting. Additionally, in the last convolution layer, the
 transformed 2D images are generated by applying only one $3\times3$
 filter.

 With the 2D images generated from individual slices, an overall 3D
 image can be obtained by stacking these 2D image slices back
 together. A limitation of such slice-based 3D transformation is that
 the contextual information between slices is overlooked and not taken
 into account to generate the overall conversion. Our strategy to
 tackle this issue is to combine the conversion results from three
 orthogonal views -- sagittal, coronal and axial. While not directly
 extracting 3D information from sub-volumes, slices from these
 orthogonal views do contain complementary information, enough to
 depict the 3D neighborhood surrounding each voxel.

    

 FCN/U-Net segmentation can also be carried out with 3D convolutions
 \cite{3D-U-Net}, but we believe our ``2D U-Convert-Net+ multi view''
 have several practical advantages, especially for whole-brain
 neuroimage analyses.


 \subsection{Loss Functions and Evaluation Metrics}

 As mentioned in last section, the objective to be optimized in our
 U-Convert-Net is the mean squared error (MSE) between the generated
 3T images and ground truth.  This is also a very common setup in SR
 algorithms, as minimizing MSE also maximizes the peak signal-to-noise
 ratio (PSNR), which is a well-accepted measure used to evaluate and
 compare SR algorithms \cite{zhang2016colorful}. Structural similarity
 index (SSIM) is another widely used evaluation metric in SR
 studies. PSNR is good at showing absolute errors, while SSIM focuses
 more on structural resemblance which carry important information on
 human's visual scene.

 However, minimizing MSE tends to produce blurry results
 \cite{pathak2016context, zhang2016colorful}. This is because the
 minimum of MSE would be the average of all possible samples, which
 leads to blurring. Identifying loss functions that can force FCN-like
 models to produce sharp, realistic images, is still an open problem
 and may have to be application-dependent.

GAN models are designed, to certain extent, to tackle this
issue. GANs, including SRGAN, learn a loss that trains the
discriminator to be intelligent at identifying if the output image is
real or fake, while simultaneously train the generator to minimize this
loss. Blurry images will be rejected as they do not look real.
The perceptual loss adopted in SRGAN 
consists of an adversarial loss and a content loss.
Minimizing the content loss would increase the similarity,
perception-wise instead of pixel-wise, between the generated images
and the target.

The enhanced conversion quality obtained in SRGAN, however, is not
reflected in the evaluations with traditional metrics -- SGRAN is
outperformed by its MSE-based version measured with PSNR.  The authors
of SRGAN had to resort to a subjective metric called {\it mean opinion
  score} (MOS) to demonstrate the improvements. In MOS, human raters
are called in to assign integral scores to assess the obtained
results.

We are well aware of the above issues, and take them into account in
our experiment designs for MR conversion, as well as the analysis of
the results.

\section{Experiments}

\subsection{Data, Preprocessing, and Evaluation Metrics}
The data used in this work were obtained from the ADNI database
\cite{ADNI}.
Started in 2004, ADNI is
currently in the stage of ADNI 3, after the completion of ADNI 1, GO and
2. The participants enrolled in late stages (GO/2/3) were scanned
using the 3T MRI scanning
protocols. In ADNI 1, however, subjects were
scanned using either a 1.5T scanner, or both 1.5T and 3T scanners. MR
images acquired in the latter generates a considerable number of
matched 1.5T/3T pairs (in the sense of same person in the same
visit). We searched the ADNI database for these cases, and downloaded
all the 1.5T/3T pairs acquired in the same visits (one pair each
visit). To simplify our network design, we chose the largest subset of
the images that have the same resolutions/sizes, which results in 157
distinct pairs of 1.5T/3T images, from 47 different subjects.

A series of pre-processing steps were then applied. Firstly, each MRI
pair was spatially aligned using SPM12. The skulls were removed, and
the image size was uniformed reduced to ($256 \times 256 \times 100$)
after removing a number of empty slices. As the 3T images in ADNI
GO/2/3 were all acquired along the sagittal view, 
each 1.5T MR image was aligned with their 3T counterpart, and
resampled into 2D slices along the same axis. These 157 pairs of
images are the training samples in this study. In our experiments, we
randomly chose $90\%$ of subjects (totally 141) for training, and the
remaining one tenth (16) for testing.


As mentioned in the previous section, the similarity between the
generated 3T images and ground truth are measured using SSIM and
PSNR. We also qualitatively look into the image contrast at various
areas, including the boundaries between gray matter and white matter.

\subsection{Results}


All the experiments were conducted on an NVidia GTX 1080 GPU (2560
CUDA Cores, 8 GB GDDR5X Memory) using TensorFlow package. We design a
number of experiments to evaluate our model, as well as the competing
solutions. The tests are on running-time and conversion accuracy,
respectively.

{\bf Experimental design} Through early experiments, we observed that
the four models had rather different performance in terms of running
time. PRSR was very slow, and would take a prohibitively long time to
process the original $256 \times 256$ sized images. Therefore, we
design the first experiment to specifically evaluate the time
performance of the models. In order to ensure PRSR to run through, we
reduce the image size to $96 \times 96$ in this experiment. The results will
be reported shortly. The other three models can take the $256 \times 256$
sized images, though still with rather different GPU times. The second
experiment is designed to evaluate the accuracies of modality
conversion of the models. Training and testing are conducted on the
original image pairs, with PRSR being excluded for comparison. The
third experiment is to evaluate the effectiveness of multi-view
ensemble, when integrated on top of U-Convert-Net.

The training details of the first two experiments, including image
sizes, learning rates, training time and testing time, are listed in
Table 1. In experiment 3, the same settings have been used for the
U-Convert-Nets on coronal and axial views. All the experiments were
run over 40 epochs.

 \begin{table}[!ht]
\small
\centering
\caption{\small Training details of the experiments: image sizes,
  learning rates (lr), batch sizes, training time (seconds per epoch)
  and testing time (seconds per image).}
\vspace{0.05in}
\scalebox{0.85}{
\begin{tabular}{c|c|c|c|c|c}
\hline
\hline
\multicolumn{1}{c|}{image size} &
\multicolumn{1}{c|}{...} &
\multicolumn{1}{c|}{PRSR} &
\multicolumn{1}{c|}{SRGAN} &
\multicolumn{1}{c|}{ESPCN} &
\multicolumn{1}{c}{U-Convert-Net} \\
\hline

\multirow{4}{1.1cm}{\centering exp1 96$\times$96} & lr & 0.004 & 0.001 &0.002 &0.001 \\
                                                  \cline{2-6}
                                                  & batch & 4 & 4 &4 &4 \\
                                                  \cline{2-6}
                                                  & training & 1666 & 1998 & 92 & 203 \\
                                                  \cline{2-6}
                                                  & testing & 1921 & 0.005 &0.005 &0.005 \\
\hline                

\multirow{4}{1.1cm}{\centering exp2 256$\times$256} & lr & -- & 0.001 &0.001 &0.001 \\
                                                    \cline{2-6}
                                                    & batch & -- & 1 &4 &4 \\
                                                    \cline{2-6}
                                                    & traing & -- & 8979 & 118 &313 \\
                                                    \cline{2-6}
                                                    & testing & -- & 0.005 &0.005 &0.005 \\
 
\hline
\hline
\end{tabular}
}
\vspace{-0.5mm}
\label{T:settings}
\end{table}

{\bf Running time comparison} 
The results are shown in Table~\ref{T:settings}. For the tests on
$96 \times 96$ sized images, ESPCN is the fastest, followed by
U-Convert-Net, PRSR and SRGAN.
PRSR takes much longer time in testing, as pixels need to
be generated one-by-one in a serial order. In the tests for the
full-sized images (experiment 2), ESPCN is still the fastest.  This is
due to the efficient design in ESPCN that specifically shortcuts the
deconvolution operations. Our U-Convert-Net is slower than ESPCN, but
is 28 times faster than SRGAN. This can be partially explained by the
fact that SRGAN employs ResNet, with much more layers than
U-Convert-Net.

 \begin{table}[!ht]
\small
\centering
\caption{\small System performance measured with SSIM and PSNR.}
\vspace{0.05in}
\scalebox{0.8}{
\begin{tabular}{c|c|c|c|c}
\hline
\hline
\multicolumn{1}{c|}{.....} &
\multicolumn{1}{c|}{ESPCN} &
\multicolumn{1}{c|}{SRGAN} &
\multicolumn{1}{c|}{U-Convert-Net} &
\multicolumn{1}{c}{U-Convert-Net+3Vs} \\
\hline
 SSIM          & 0.85  & 0.94 & 0.93 & 0.94 \\ 
  \hline
 PSNR          & 21.8  & 25.1 & 26.5 & 28.4\\
  \hline
\hline
\end{tabular}
}
\vspace{1.5mm}
\vspace{-2mm}
\label{T:measure}
\end{table}

{\bf Conversion Accuracy by Traditional Metrics} The quantitative
evaluations of the solutions, measured with SSIM and PSNR, is shown in
Table~\ref{T:measure}.
All the measurements are conducted on the 2D slices extracted from the
sagittal view. It can be observed that U-Convert-Net achieves better
performance than ESPCN and SRGAN in terms of PSNR. For SSIM,
U-Convert-Net is better than ESPCN and is comparable with SRGAN but
with a much shorter training time. The effectiveness of multi-view
ensemble is validated in experiment 3: in addition to the original
sagittal-view slices/results, we conduct the network constructions and
trainings for the slices along coronal and axial views. The 3T
conversion volumes of the three views are averaged to produce the
final output of our multi-view U-Convert-Net. As shown in the last
column of Table~\ref{T:measure}, both SSIM and PSNR are improved
through the view ensemble.

\begin{figure}  
  \begin{tabular}{ccc}
 \includegraphics[width=0.23\columnwidth]{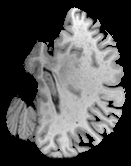} & 
 \includegraphics[width=0.23\columnwidth]{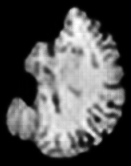} & 
 \includegraphics[width=0.23\columnwidth]{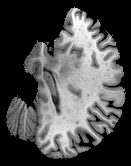} \\
 
  \small{1.5T MR image} & \small{ESPCN} & \small{SRGAN}  \\
  
 \includegraphics[width=0.23\columnwidth]{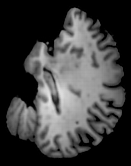} &
 \includegraphics[width=0.23\columnwidth]{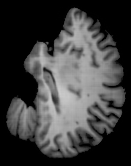} &
 \includegraphics[width=0.23\columnwidth]{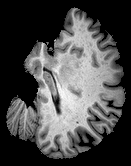}  \\

 \small{U-Convert-Net} & \small{U-Convert-Net$+$mult-view} & \small{3T MR image} \ \
\end{tabular}
 \vspace{0.15in}
 \caption{\small Source (1.5T), target (3T) and the generated 3T MR
   images using different models. Figures are best viewed on screen. }
 \label{fig:gan_fcn}
 \end{figure}

 {\bf Actual Contrast Enhancements} A major advantage of 3T MRIs over
 1.5T is the enhanced contrasts along the boundaries of gray and white
 matter.  To get a qualitative comparison, we also look into the 3T
 images produced by the models. Fig.~\ref{fig:gan_fcn} shows the
 zoom-in version of a typical slice. Obviously ESPCN's high speed
 comes with a cost -- its 3T result is over smoothed with missing
 details. Results from SRGAN, U-Convert-Net and its multi-view version
 are all consistent with the ground-truth 3T image.

 The comparison between multi-view U-Convert-Net and SRGAN reveals the
 deficiencies of the traditional metrics. While the former has a
 ``superior'' performance measured by PSNR and SSIM, SRGAN obviously
 generates slices that have higher contrasts between gray and white
 matter, as shown in Fig.~\ref{fig:gan_fcn}. This observation is
 consistent with the experiments conducted in SRGAN paper
 \cite{Jun-Yan2017Photo}, where the proposed method (SRGAN) gets lower
 PSNR and SSIM measures than its MSE version, but performs better in
 the subjective metric MOS, judged by human raters.

 All in all, our observations can be summarized as: 1) SRGAN can
 produce higher contrasts MR images with the help of its
 discriminator; 2) the multi-view U-Convert-Net framework produces
 results with higher PSNR and SSIM measures; 3) multi-view
 U-Convert-Net framework can be trained much faster than SRGAN.

\section{Conclusions}

In this work, we study four solutions for MRI modality conversion. The
models are compared and evaluated with various metrics. The FCN-based
Multi-view U-Convert-Net is fast to train and it achieves the better
performance in 1.5 to 3T MRI mapping, measured with SSIM and
PSNR. SRGAN, on the other hand, produces more realistic 3T-like images
with enhanced contrasts. The take-home messages could be: 1) GAN
models have inherent mechanism for image conversions, including that
between MRIs; 2) FCNs equipped with other loss functions, including
perception-based losses, are also worth explorations for faster
solutions.

%


\bibliographystyle{IEEEtranS}
\bibliography{liu_icpr18, new_refs}

\end{document}